# Acoustic Steering of Active Spherical Carriers

**Majid Rajabi***

**Assistant Professor*
*Sustainable Manufacturing Systems Research Laboratory*
*School of Mechanical Engineering, Iran University of Science and Technology,*
*Narmak, Tehran, Iran.*
*Phone : +98-21-77240649*
**Email of Corresponding Author:* majid_rajabi@iust.ac.ir

**Hossein Khodavirdi**

*Sustainable Manufacturing Systems Research Laboratory*
*School of Mechanical Engineering, Iran University of Science and Technology,*
*Narmak, Tehran, Iran.*

**AlirezaMojahed**

*Linear and Nonlinear Dynamics and Vibrations*
*Laboratory, Department of Mechanical Science and Engineering,*
*University of Illinois at Urbana-Champaign, Urbana, Illinois 61801-*
*2307, USA.*

## Abstract

Following the novel introduced concept of the active carriers, this paper brings forward a technique toward the manipulability of an internally piezo-equipped active spherical carrier subjected to the progressive acoustic plane waves as the handling contactless asset. It is assumed that the piezoelectric part of the active carrier may be actuated as a bi-sectional body (i.e., two continuous hemispherical parts), with prescribed phase difference, and the polar position of the imaginary separating plane may be altered. This issue brings about an asymmetry in the dynamics of the problem which leads to emergence of position/frequency dependent acoustic radiation torque. It is obtained that as the carrier is excited by imposing harmonic voltage with the same amplitude and a π-radians phase difference, the zero-radiation force situation is obtained for a specific amplitude and phase of voltage as a function of frequency. This situation is treated as a criterion to determine the optimal amplitude of operation voltage. It is shown that the net force's direction exerted on the activated carrier may be steered along any desired orientation, assuming the fixed direction of incident wave field. The explained method of excitation and controllability of the spatial position of the divisor plane can possibly be a breakthrough in acoustic handling of active carriers. Noticeably, by this new technique, the single beam acoustic based contact-free methods and their applications in association with the concept of active carriers are now one step forward.

## 1. Introduction

The efforts on non-contact methods for manipulations[1-4] , levitation [5], sorting [6] and focusing [5,7-10] of small bodies have inaugurated serious developments in diverse areas as medicine (e.g. drug delivery systems, cell therapies and contrast agents), engineering (e.g. material/agent transportation systems, cleaning, etc) and Lab on a chip applications (e.g., separation, sorting and patterning).

For all the methods in acoustic handling, force as a major concept to induce motion in a desired path is required. The so-called Acoustic Radiation Force (ARF) should be generated in all the polar directions in3D (three dimensional space) to assure that a complete manipulation is theoretically achievable. Many researchers devoted their research on the features and aspects of ARF, considering the magnitude evaluation [1,4], object material and geometry effects [11-15], etc. As the first step toward 1-D manipulation, the generation of negative radiation force has attracted a lot of attentions via introducing various beam forms of cross beams [9,10,16,17], Bessel [12,18-27], Gaussian [28] and Mathieu [29].

Considering the challenges of complex beam form generation, as an alternative or further proposal, the novel concept of acoustically activated bodies as self-propulsive carriers [30,31], self-activated carriers with handling incident beam [32-36], or externally activated bodies with compound fields [37,38] have been introduced in literature. It is shown that the distortion of wave field may lead to generation of interesting negative radiation force or pulling effects in association with the positive radiation force or pushing effects. Considering the directional cover by external transducer or incident wave beam in addition to the forward and backward manipulation of object along the propagation direction of the beam, the 3D translational manipulation is achievable. Here, it is assumed that the direction of incident beam is fixed and the raised question is as follows: Is it possible to internally activate the body/carrier so that the full 3D translational manipulation occurs? May the body/carrier move perpendicular to the beam direction?

In the present work, with the hope of generating force in a full angular range about the spherical object for a full directional manipulation of the object, a new method is introduced. This method takes advantage of a bi-sectional piezoelectric internal part as the driver of the active carrier which the direction of its characteristic divisor plane may be altered. This paper is just an analytical feasibility study of the concept of the possibility of full directional steering and the technical issues on practical implementation of the active part are not the concern.

# 2. Formulation

## 2.1. Configuration

Through this segment, the schematic configuration of the problem and also some technical details about the governing conditions of the problem should be delineated. The following active elastic sphere is located on the way of the plane harmonic progressive acoustic wave. The medium of the wave-sphere system is an ideal and non-viscous acoustic fluid.

As Fig. 1 displays, the elastic casing of the sphere with the farthest distance of *a*from the center of the sphere is bonded to a two-section radially polarized piezoelectric layer with inner and outer radii *c* and *b*, respectively. The two different parts of the inner hollow-sphere piezoelectric are supplied in a way which they have a phase difference of π, but both have a same voltage amplitude in each frequency.

The global coordinate system ( $X$ , $Y$ , $Z$ ) is located at the center of the sphere so that its $Z$-direction is aligned with the propagation direction of the incident wave. The local coordinate system ( $X'$ , $Y'$ , $Z'$ ) is selected so that its $Z'$ -axis is the symmetry axis of the carrier and $X'Y'-$plane is the voltage divisor plane. Also, $\theta$ and $\phi$ are selected as polar and azimuthal coordinates in the local coordinate system. Moreover, it is assumed that the position of these two piezoelectric sections (i.e., divisor plane) is maneuverable in colatitude or polar direction, $\theta$. The position of the divisor plane with respect to the incident wave propagation direction may be indicated by $\alpha$. It is obvious that any conclusion about the maneuverability of the carrier in $XY-$Plane can be expanded to 3-D space.

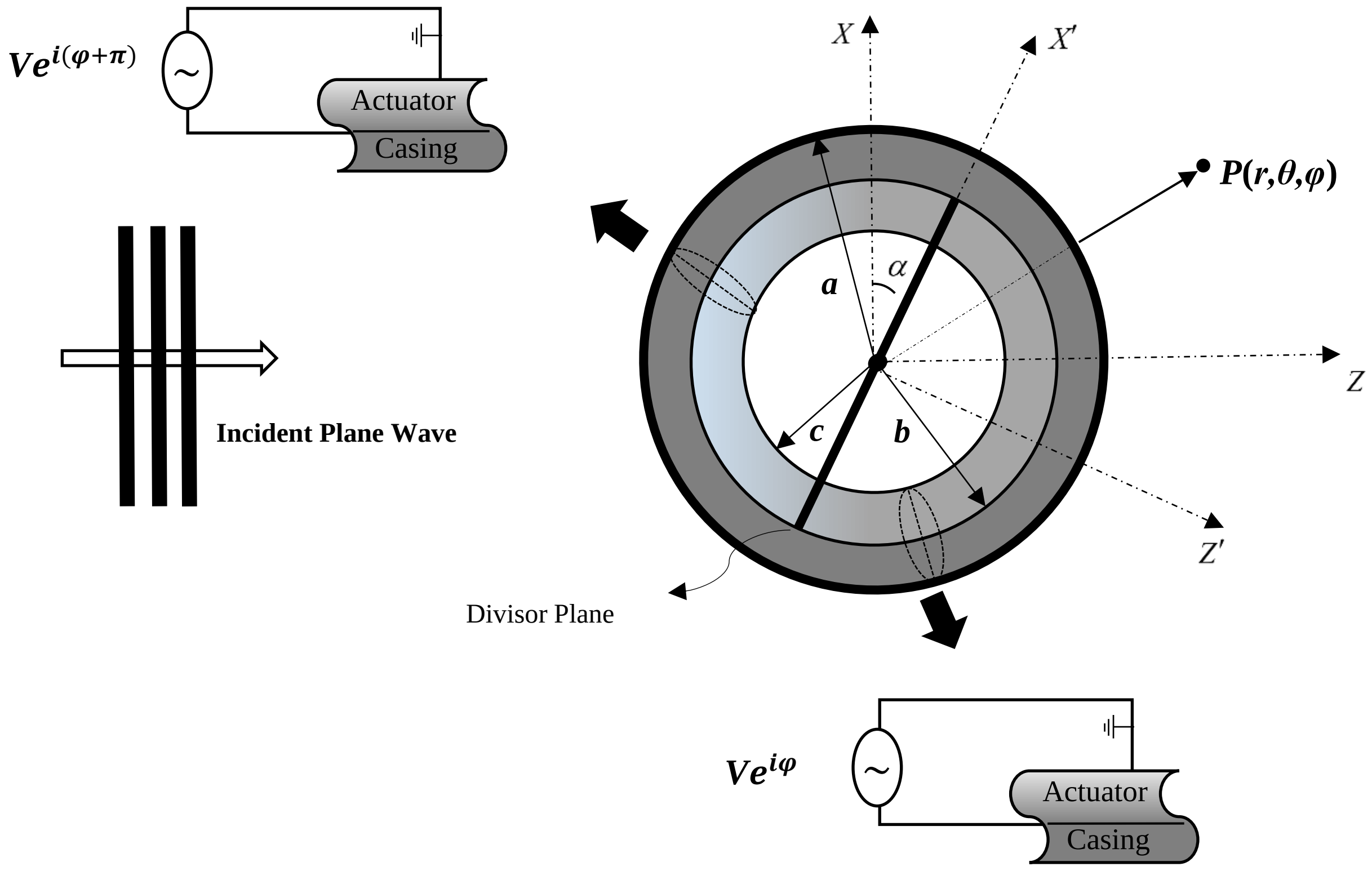

**Fig.1.**Illustration of the Problem: A plane progressive acoustic beam is heading a two-division active carrier with a phase difference of π between its two parts and a controllable plane which determines the polar position of the two piezo-parts.

### 2.2. Acoustic Field Equations

On the way of solving the problem, it was preferred to use an arbitrary oriented acoustic plane wave and change the angle of wave in mathematical models and equations to show the change in relative status between the wave and the two sides of the sphere. The velocity potential function of incident wave field based on the partial wave expansion method [39-41] in linear acoustics (i.e., $\nabla^2\varphi = -k_f^2\varphi$ where $\nabla\varphi = \mathbf{v}$ is the velocity vector field in the fluidic medium and $k_f = \omega / c_f$ is the wave number in which $c_f$ is the sound speed in the fluidic medium )may be expanded as:

$$\varphi_{inc} = \varphi_0 \sum_{n,m} i^n j_n(k_f r) Y_{nm}^*(\alpha, 0) Y_{nm}(\theta, \phi) e^{-i\omega t}, \tag{1}$$

In this equation, $\sum_{n,m}$ means $\sum_{n=0}^{\infty} \sum_{m=-n}^{+n}$ , $Y_{nm}$ is the spherical harmonic function with *m*-th degree and order *n*, $j_n(.)$ is the spherical Bessel function of the first kind of order *n* and (.)* means complex conjugate of a variable. It is also noticeable that due to the presence of term $e^{-i\omega t}$ in the both equations of wave and the normal modes expansion of variables in boundary conditions, it can be neglected for simplicity.

By considering Sommerfeld radiation condition for scattered field, the solution to the time-independent form of fundamental equation will be:

$$\varphi_{sc} = \varphi_0 \sum_{n,m} i^n A_{nm} h_n^{(1)}(k_f r) Y_{nm}^*(\alpha, 0) Y_{nm}(\theta, \phi) e^{-i\omega t}. \tag{2}$$

$h_n^{(1)}(.)$ in this equation is the spherical Hankel function of the first kind and *n*-th order and $A_{nm}$ are unknown modal scattering coefficients which will be determined by using appropriate boundary conditions. The relation between the velocity potential and pressure field for both incident and scatter field is:

$$P(r, \theta, \omega) = \rho_f \frac{\partial \varphi}{\partial t}, \tag{3}$$

where $\rho_f$ is the density of the fluidic medium. So by using Eqs. (1), (2) and (3) and also by utilizing the superposition principle in linear acoustic regime, the two following equations for total velocity potential function and acoustic pressure at arbitrary point $P(r,\theta,\phi)$ can be obtained like:

$$\varphi_{tot} = \varphi_0 \sum_{n,m} i^n Y^*_{nm}(\alpha,0) Y_{nm}(\theta,\phi)(A_{nm} h_n^{(1)}(k_f r) + j_n(k_f r)) e^{-i\omega t}, \tag{4}$$

$$p_{tot} = \omega \rho_f \varphi_0 \sum_{n,m} i^{n+1} Y^*_{nm}(\alpha,0) Y_{nm}(\theta,\phi)(A_{nm} h_n^{(1)}(k_f r) + j_n(k_f r)) e^{-i\omega t}. \tag{5}$$

### 2.3. Dynamics of Structure

Based on the previous works, the structural equations for elastic and piezoelectric layers are given as:

$$\boldsymbol{\sigma} = \mathbf{cs}, \ \boldsymbol{\Sigma} = \mathbf{CS} - \boldsymbol{\varpi}\Phi, \ \boldsymbol{\Delta} = \mathbf{eS} + \mathbf{Q}\Phi, \tag{6}$$

where $\boldsymbol{\sigma}$ and $\mathbf{s}$ indicate the stress and strain vectors of the isotropic elastic material respectively, and $\boldsymbol{\Sigma}$, $\mathbf{S}$, $\boldsymbol{\Delta}$ are the stress, strain and electric field of the piezoelectric material, respectively . Also, $\Phi$ refers to the electric potential function. The elastic constant matrices of the casing/actuator, $\mathbf{c}$ and $\mathbf{C}$, the piezoelectricity matrix, $\mathbf{e}$, and the operator matrices $\boldsymbol{\Lambda}$ and $\mathbf{Q}$ are given in [32]. The strain - displacement relations are as $\mathbf{s} = \mathbf{Hu}$, $\mathbf{S} = \mathbf{HU}$, where $\mathbf{u}$ and $\mathbf{U}$ are displacement vectors of the isotropic elastic material and the piezoelectric material, respectively and $\mathbf{H}$ is again given in [32].In the absence of the body forces, the relevant structural dynamic equations may be described as

$$\bar{\boldsymbol{\Upsilon}}\boldsymbol{\sigma} = \rho_c \ddot{\mathbf{u}}, \ \bar{\boldsymbol{\Upsilon}}\boldsymbol{\Sigma} = \rho_p \ddot{\mathbf{U}}. \tag{7}$$

where $\rho_c$ and $\rho_p$ are the densities of the casing and piezoelectric actuator materials, respectively, and the matrix, $\bar{\boldsymbol{\Upsilon}}$, is represented in [32].

Assuming the free charge density condition, the Gaussian equation of electric equilibrium is represented as $\nabla_2 \Delta_r + \Delta_r + (1/\sin\theta)\partial/\partial\theta(\Delta_\theta \sin\theta) + (1/r\sin\theta)\partial/\partial\phi(\Delta_\phi) = 0$, where $\nabla_2 = r\partial/\partial r$ .

### 2.4. Applied Voltage and Scattering Coefficients

The prescribed applied electrical voltage imposed at the inner and outer surfaces of piezo-actuator is:

$$\begin{aligned} &\Phi(r=b,\theta,\phi,t)=V(\theta,\phi,\omega)e^{-i\omega t}, \\ &\Phi(r=c,\theta,\phi,t)=0. \end{aligned} \tag{8}$$

where $V\ (\theta,\omega)$ denoted the imposed electric potential amplitude distribution on the spherical surface, indicated by Heaviside function as:

$$V(\theta,\phi,\omega)=\begin{cases} 0\le\theta<\dfrac{\pi}{2}, \rightarrow +V, \\ \dfrac{\pi}{2}\le\theta\le\pi, \rightarrow -V, \end{cases} \tag{9}$$

The voltage can be expanded in its non-dimensional form as

$$V(\theta,\phi,\omega)=\frac{ae_{33}}{\varepsilon_{33}}\sum_{n,m}\Phi_n^m(r=b)Y_{nm}(\theta,\phi), \tag{10}$$

Putting into practice the orthogonality for Eq. (10), along with using Eq. (9) will bring about electric potential $\Phi_n$. So first we do some manipulations to use orthogonality like:

$$-VY_{n,m}^*(\theta,\phi)\sin\theta=-\frac{ae_{33}}{\varepsilon_{33}}\sum_{n,m}\Phi_n^m(r=b)Y_{nm}(\theta,\phi)Y_{n,m}^*(\Omega,0)\sin\theta, \tag{11}$$

$$\Phi_n^m=\frac{\varepsilon_{33}}{ae_{33}}\int_0^{\pi}\int_0^{2\pi}VY_{n,m}^*(\theta,\phi)\sin\theta d\theta d\phi, \tag{12}$$

Also we know that,

$$Y_{n,m}^*(\theta,\phi)=(-1)^m Y_{n,-m}^*(\theta,\phi), \tag{13}$$

where,

$$Y_{nm}(\theta,\phi)=\sqrt{\frac{(2n+1)(n-m)!}{4\pi(n+m)!}}P_n^m(\cos\theta)e^{im\phi}, \tag{14}$$

and,

$$P_n^{-m} = (-1)^m \frac{(n-m)!}{(n+m)!}, \tag{15}$$

so we get,

$$\Phi_n^m = \frac{\varepsilon_{33}}{ae_{33}} \underbrace{\int_0^{2\pi} e^{-im\phi} d\phi}_{0 \xrightarrow{if} m\neq 0} \int_0^{\pi} V P_n^m(\cos\theta)\sin\theta d\theta, \tag{16}$$

Now we set m=0 and work on the second integral by using the expression for integration of Legendre polynomials which is:

$$\int P_n(x)dx = \frac{1}{2n+1}\left[d\left(P_{n+1}(x) - P_{n-1}(x)\right)\right], \tag{17}$$

Eventually, the expression for $\Phi_n$ would be,

$$\Phi_n = \frac{\varepsilon_{33}}{ae_{33}} V\sqrt{\pi(2n+1)}F_n, \tag{18}$$

where $\varepsilon_{33}$ is the dielectric constant, $e_{33}$ is the piezoelectric constant and $F_n$ is defined as:

$$F_n = \left\{\frac{1}{2n+1}\left[p_{n+1}(0) - p_{n-1}(0) + p_{n+1}(-1) - p_{n-1}(-1) - 2p_{n+1}(0) + 2p_{n-1}(0)\right]\right\}. \tag{19}$$

and $p_n^m(x)$ is the associated Legendre polynomials. The counter *m* for the obtained expression of electric potential is omitted due to the integral operation on the period of [0,2π] on $e^{im\phi}$ which we have to set $m=0$ to achieve non-zero values for the electric potential, $\Phi_n$.By using the separation of variables method and following the so-called State-Space solution, some mathematical actions can be taken to find the relationships between the state vectors at *r* equals to *c*, *b* and *a*, via the global transfer matrices. Applying the appropriate boundary conditions like what we have done in Ref. [32] at the outer and inner surfaces of the structure leads to the scattering coefficients, $A_{nm}$, which uncover all the physical quantities in the fluidic medium. To be more specific, by using Eqs. (6) and (7) and by considering the equation for Gaussian electric equilibrium a change of variables can be used to obtain the solution to the decoupled equations of motion. Eventually, after imposing the boundary conditions and doing some mathematical manipulations the following equation is obtained:

$$\Omega_n^m X_n^m = T_n^m, \tag{20}$$

where $\Omega_n^m$, $X_n^m$ and $T_n^m$ are explained in the appendix. Noticeably, there is no demand to re-write the process of finding the scattered coefficients as we have done it in detail in one of our previous works [32].

**2.5. Acoustic Radiation Force and Torque**

Considering the nonlinear acoustic regime, it is expected for an object affected by a sound wave field to experience the so-called acoustic radiation force (i.e., $\langle \mathbf{F} \rangle = \left\langle \iint_{\Upsilon_0} \mathbf{\Pi}.\mathrm{d}\mathbf{\Upsilon} \right\rangle$, where $\mathbf{\Upsilon}_0$ is the fixed outer surface of the body at equilibrium, $<.>$ means time averaged over a cycle of oscillations, the Brillouin radiation stress tensor is defined as $\Pi_{ij} = -< p > \delta_{ij} - \rho < v_i v_j >$, in which $\delta_{ij}$ is Kronecker delta, $\rho$ is the density of ambient medium and $v_i$ are the particle velocity components in the surrounding fluid medium) and torque (i.e., $\langle \mathbf{N} \rangle = \left\langle \iint_{\Upsilon_0} \mathbf{r} \times (\mathbf{\Pi}.\mathbf{n}) \mathrm{d}\mathbf{\Upsilon} \right\rangle$, where $\mathbf{r}$ is the position vector and $\mathbf{n}$ is the normal unit vector of the surface, [42]).

The general form of the time-averaged radiation force in both *X* and *Z* directions can be written as $F_j = E_{inc} S_c \mathit{\Gamma}_j$ where $j = X, Z$, $S_c = \pi a^2$ is the cross-sectional area of the object, $E_{inc} = \rho_f k_f^2 \varphi_0^2 / 2$ is a representative of incident wave energy density and $\mathit{\Gamma}_j$ is the non-dimensional acoustic radiation force in *j*-direction. The radiation torque along *Y*-axis can be written as $N_y = \pi a^3 E_{inc.} \tau_y$, where $\tau_y$ is the non-dimensional radiation torque. Now, by using what is indicated in [42,43] about radiation force and torque formula and by adjusting the parameters in formula for this case, we have

$$\Gamma_X = \frac{1}{2\pi (k_f a)^2} \operatorname{Im} \sum_{n,m} (a_n^m + d_n^m)(d_{n+1}^{m-1,*} b_{n+1}^{-m} + d_{n-1}^{m-1,*} b_n^{m-1} - d_{n+1}^{m+1,*} b_{n+1}^{m} - d_{n-1}^{m+1,*} b_n^{-m-1}), \tag{21}$$

$$\Gamma_Z = \frac{1}{\pi (k_f a)^2} \operatorname{Im} \sum_{n,m} (a_n^m + d_n^m)(d_{n+1}^{m,*} c_{n+1}^{m} + d_{n-1}^{m,*} c_n^{m}), \tag{22}$$

$$\tau_y = -\frac{1}{2\pi (k_f a)^3} \operatorname{Im} \sum_{n,m} (a_n^m + d_n^m)(d_n^{m+1,*} e_n^{m} - d_n^{m-1,*} e_n^{-m}), \tag{23}$$

where,

$a_n^m = 4\pi i^n Y_{nm}^*(\alpha,0)$, $b_n^m = \sqrt{\left[(n+m)(n+m+1)\right]/\left[(2n-1)(2n+1)\right]}$,

$c_n^m = \sqrt{\left[(n-m)(n+m)\right]/\left[(2n-1)(2n+1)\right]}$, $d_n^m = 4\pi i^n Y_{nm}^*(\alpha,0) A_n^m$, $e_n^m = \sqrt{(n+m)(n+m+1)}$.

The direction of manipulation may be indicated by $\gamma = \tan^{-1}(\Gamma_X / \Gamma_Z)$ as the angle between the net radiation force direction and Z-axis.

## 2. Acoustic Manipulation Technique

The acoustic manipulation can be accomplished perfectly, if the produced acoustic radiation force can cover the whole angular range in *XZ*- plane. Moreover, this concept makes sense when we are able to produce a stable acoustic radiation torque in *Y*-direction and in addition, the special cases of mere upward-downward forces and mere attractive-repulsive forces are achievable.

For the earliest step toward the manipulation technique, the range of the amplitude of the applied voltage should be determined. Considering the previous works in this area [32], and limiting the problem to the symmetric configuration (i.e., the divisor plane is perpendicular to the incident wave propagation), it is expected that for any operational frequency, a specific voltage amplitude may be found that determines the radiation force state on the carrier (i.e., negative, positive and zero radiation forces).

Now, the problem should be described mathematically to determine the required voltage amplitude to produce the force in all the desired directions by just adjusting the angle of the divisor plane. By considering that the effect of non-uniform imposed voltage is present in all the vibrating modes, the procedure of finding the cancellation voltage line in $\bar{\varsigma}\bar{\eta}-$plane gets a little longer where $\bar{\varsigma}$ and $\bar{\eta}$ are real and imaginary parts of the voltage, respectively (i.e., $V=\bar{\varsigma}+i\bar{\eta}$).

Knowing the expression for the scattering coefficients in terms of the electrical potential from [32] yields

$$A_n^m = Z_1^{n,m} + Z_2^{n,m}\,\Phi_n, \tag{24}$$

where $Z_1^{n,m}$ and $Z_2^{n,m}$ are indicated in appendix which it is obtained by using Cramer rule on equation 20. Moreover, the scattering coefficients can be written as

$$A_n^m = \varsigma_n^m + i\eta_n^m, \tag{25}$$

where $\varsigma_n^m$ and $\eta_n^m$ are real and imaginary parts of scattered coefficients. Having in mind that all the body modes play a role in the generation of acoustic radiation force, using Eqs. (24) and (25), and by using the formula of acoustic radiation force in *Z*-direction and in the conditions that the wave is perpendicular to the divisor plane, we get

$$\varsigma_m^n = \mathrm{Re}(Z_1^{n,m}) + \left\{\frac{\varepsilon_{33}}{ae_{33}}F_n\sqrt{\pi(2n+1)}\left[\bar{\varsigma}\,\mathrm{Re}(Z_2^{n,m}) - \bar{\eta}\,\mathrm{Im}(Z_2^{n,m})\right]\right\}, \tag{26}$$

$$\eta_m^n = \mathrm{Im}(Z_1^{n,m}) + \left\{\frac{\varepsilon_{33}}{ae_{33}}F_n\sqrt{\pi(2n+1)}\left[\bar{\varsigma}\,\mathrm{Im}(Z_2^{n,m}) - \bar{\eta}\,\mathrm{Re}(Z_2^{n,m})\right]\right\}, \tag{27}$$

and then we can rewrite the zero state of acoustic radiation force in *Z*-direction (i.e., $\Gamma_Z = 0$) as:

$$A\bar{\varsigma} + B\bar{\eta} + C = 0, \tag{28}$$

which *A*, *B* and *C* are obtained after some manipulations as

$$\begin{aligned} A = \frac{-4}{(k_f a)^2} \sum_{n=0}^{\infty} \{ (n+1)(\frac{\varepsilon_{33}}{a e_{33}}) [ & F_n \sqrt{\pi(2n+1)} \operatorname{Re}(Z_2^{n,m}) + F_{n+1} \sqrt{\pi(2n+2)} \operatorname{Re}(Z_2^{n+1,m}) + \\ & 2\operatorname{Re}(Z_1^{n,m}) F_{n+1} \sqrt{\pi(2n+2)} \operatorname{Re}(Z_2^{n+1,m}) + 2\operatorname{Re}(Z_1^{n+1,m}) F_n \sqrt{\pi(2n+1)} \operatorname{Re}(Z_2^{n,m}) + \\ & 2\operatorname{Im}(Z_1^{n,m}) F_{n+1} \sqrt{\pi(2n+2)} \operatorname{Im}(Z_2^{n+1,m}) + 2\operatorname{Im}(Z_1^{n+1,m}) F_n \sqrt{\pi(2n+1)} \operatorname{Im}(Z_2^{n,m}) ] \}, \end{aligned} \tag{29}$$

$$\begin{aligned} B = \frac{-4}{(k_f a)^2} \sum_{n=0}^{\infty} \{ (n+1)(\frac{\varepsilon_{33}}{a e_{33}}) [ & -F_n \sqrt{\pi(2n+1)} \operatorname{Im}(Z_2^{n,m}) - F_{n+1} \sqrt{\pi(2n+2)} \operatorname{Im}(Z_2^{n+1,m}) - \\ & 2\operatorname{Re}(Z_1^{n,m}) F_{n+1} \sqrt{\pi(2n+2)} \operatorname{Im}(Z_2^{n+1,m}) - 2\operatorname{Re}(Z_1^{n+1,m}) F_n \sqrt{\pi(2n+1)} \operatorname{Im}(Z_2^{n,m}) + \\ & 2\operatorname{Im}(Z_1^{n,m}) F_{n+1} \sqrt{\pi(2n+2)} \operatorname{Re}(Z_2^{n+1,m}) + 2\operatorname{Im}(Z_1^{n+1,m}) F_n \sqrt{\pi(2n+1)} \operatorname{Re}(Z_2^{n,m}) ] \}, \end{aligned} \tag{30}$$

$$C = \frac{-4}{(k_f a)^2} \sum_{n=0}^{\infty} (n+1) [\operatorname{Re}(Z_1^{n,m}) + \operatorname{Re}(Z_1^{n+1,m}) + 2\operatorname{Re}(Z_1^{n,m}) \operatorname{Re}(Z_1^{n+1,m}) + 2\operatorname{Im}(Z_1^{n,m}) \operatorname{Im}(Z_1^{n+1,m})]. \tag{31}$$

As was already indicated in this section, we are considering the situation that the divisor plane is in its initial position in a way that the wave direction is perpendicular to the plane sothe problem remains symmetric and $Z_i^{n,m}$ are limited to $Z_i^{n,m} = 0,\ m \neq 0$. Equation (28) indicates that in each frequency, cancellation voltage is a line in $\bar{\varsigma}\bar{\eta}-$plane and the minimum cancellation voltage for each frequency would be the radius of the tangent circle which is centered at (0,0) and it is $|C| / \sqrt{A^2 + B^2}$ . Moreover, the phase of minimum cancellation voltage is equal to $\tan^{-1}(\bar{\eta} / \bar{\varsigma})$.

## 4. Numerical Results and Discussions

In order to make the mathematical expressions more perceptible, we considered a numerical example and evaluated it by a programmed code in $MATLAB^{®}$. The non-dimensional frequency range is set as $0.1 \le ka \le 50$ which stems from the practical frequency of ultrasonic transducers for $a$=1mm that is the outer radius of the spherical object. Now for convergence, the maximum truncation constant is set as $n = 70$. Moreover, the polar angle of the divisor plane is changed clock-wise and other parameters like $c_f$ (the velocity of sound wave in the water), $\rho_f$ (the density of water), maximum number of sublayers of laminar structure of the body, piezoelectric constants, Young Modulus are set as exactly what has been done in [32]. So we can validate our code in the absence of the voltage as follows:

- The figure of Non-dimensional force in Z-direction Vs. Non-Dimensional Frequency showed an exact agreement with Fig.3 in [32].
- The figures of scattering coefficients in zero, first and second modes in the absence of voltage and in the initial position of divisor plane (i.e., $\alpha = 0$) showed an exact agreement with Figs.5(a),5(b) and 5(c) in [32].

The plots are not shown in order to make the article brief and avoid unnecessary figures.

Turning to the voltage-applying process and specific voltage distribution of this problem on the piezo-actuator, needs to achieve the minimum amplitude of normalized voltage to cancel the acoustic radiation force to be able to determine the order of the operational voltage amplitude for practical acoustic handling of our active carrier. As it was already mentioned, in the symmetric configuration (i.e., $\alpha = 0$), the zero-force condition brings about a line in $\bar{\varsigma}\bar{\eta}-$ plane, as shown in Figs. 2(a) and (b), for selected frequencies of $k_f a = 2$ and $k_f a = 20$, respectively.

Figs.3 (a) and (b) show the minimum required voltage amplitude and phase, to cancel the acoustic radiation force in *Z*- direction as a function of non-dimensional frequency, $k_f a$. In Fig 3(a), the voltage is compared with the minimum required voltage for the case of breathing mode control configuration in Ref. [32]. It is seen that for great range of frequencies, the required voltage in the present configuration is generally lower than the breathing mode configuration of Ref. [32]: 1 to 2 orders for medium frequency ranges, $1 < k_f a < 10$, and 2 to 3 orders for high frequency ranges, $10 < k_f a$. At low frequency ranges, $k_f a < 1$, the order of amplitude is similar. It is due to this fact the present configuration of pseudo bi-polar excitation, the distortion of the external pressure field is more achievable than the breathing mode excitation case and the radiation force alteration is a lower effort task.

Now, the practical voltage amplitude for each operational frequency can be chosen as $R|V_{\min}|$ where $R > 1$ represents a coefficient that it should be chosen in a way that the voltage amplitude be practical and capable of producing positive, negative and zero radiation force by just changing the phase difference between the incident wave field and the applied voltage. In this case, $R = 2$ was chosen, but all the results were evaluated for some other coefficients to make sure that the executability of the handling is not dependent on a specific number. Apparently, by considering the amplitude of incident pressure field and the selected radius of the spherical object, the range of the amplitude of the voltage ($V = (a|p_{inc}|\overline{V}) / e_{33}$) is $10^{-4} \le V \le 10^{-2}$ volt which it seems feasible and practical.

Figure 4(a) shows the dependency of the radiation force on the active carrier in its symmetric configuration (i.e., $\alpha = 0$) as a function of phase difference between the incident wave field, $\tan^{-1}(\overline{\eta} / \overline{\varsigma})$, which is equivalent to the position of carrier with respect to the incident wave transducer, $z / \lambda$, where $\lambda = 2\pi c / \omega$ is the wavelength of the incident wave field.

Figure 4(b) illustrates the Nyquist plot of voltage for $R = 2$. In both figures, the stable and unstable equilibrium states are designated with *S* and *U*, respectively. It is clear that the change of position on the solid circle means change of voltage phase or the phase difference between the voltage and the incident wave field. Noticeably, for the following figures, the selected operational non-dimensional frequency is $k_f a = 2$ which it is practical and adequate regarding the radius of the sphere (i.e., $k_f a = 2$ means $\lambda / a = \pi$, which for $a = 1mm$, the operational frequency is found $\omega / (2\pi) = k_f a c_f / (2\pi a) = 470\,kHz$). As this figure shows, the zero- radiation force state occurs at two phase differences or two positions along one snapshot of the wavelength. One of them is stable equilibrium state and another is unstable. Apparently, if the active carrier is faced with the incident field, it moves toward the stable equilibrium state with translational offset below one wavelength. If we can show that the carrier is rotationally stable at this stable zero-force state, the strategy of manipulation would be similar to the standing wave method. The stable zero-force state looks like a pressure node on which the body is stable trapped and may move back and forth if the phase of the transducer is changed. Therefore, the precise handling will be achieved. The proof of rotational stability will be discussed further.

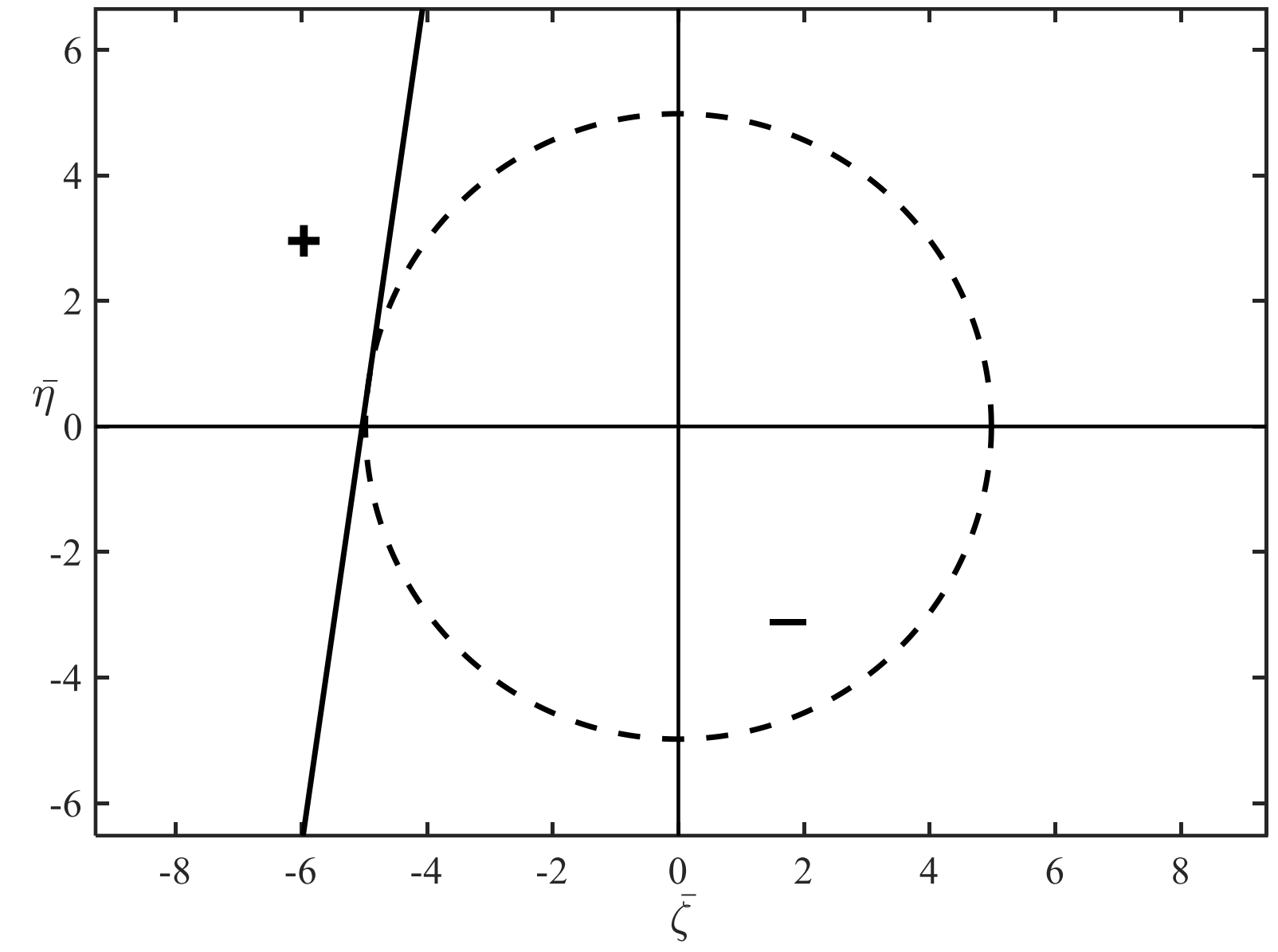

**Fig. 2(a).** Real and Imaginary parts of normalized voltage applied on the spherical object and the repulsive (positive,+) and attractive (negative,-) force zones for non-dimensional frequency of $k_f a = 2$. The black line shows the zero-radiation force state and the dashed circle shows the tangent circle, which its radius shows the minimum required voltage for zero-force state.

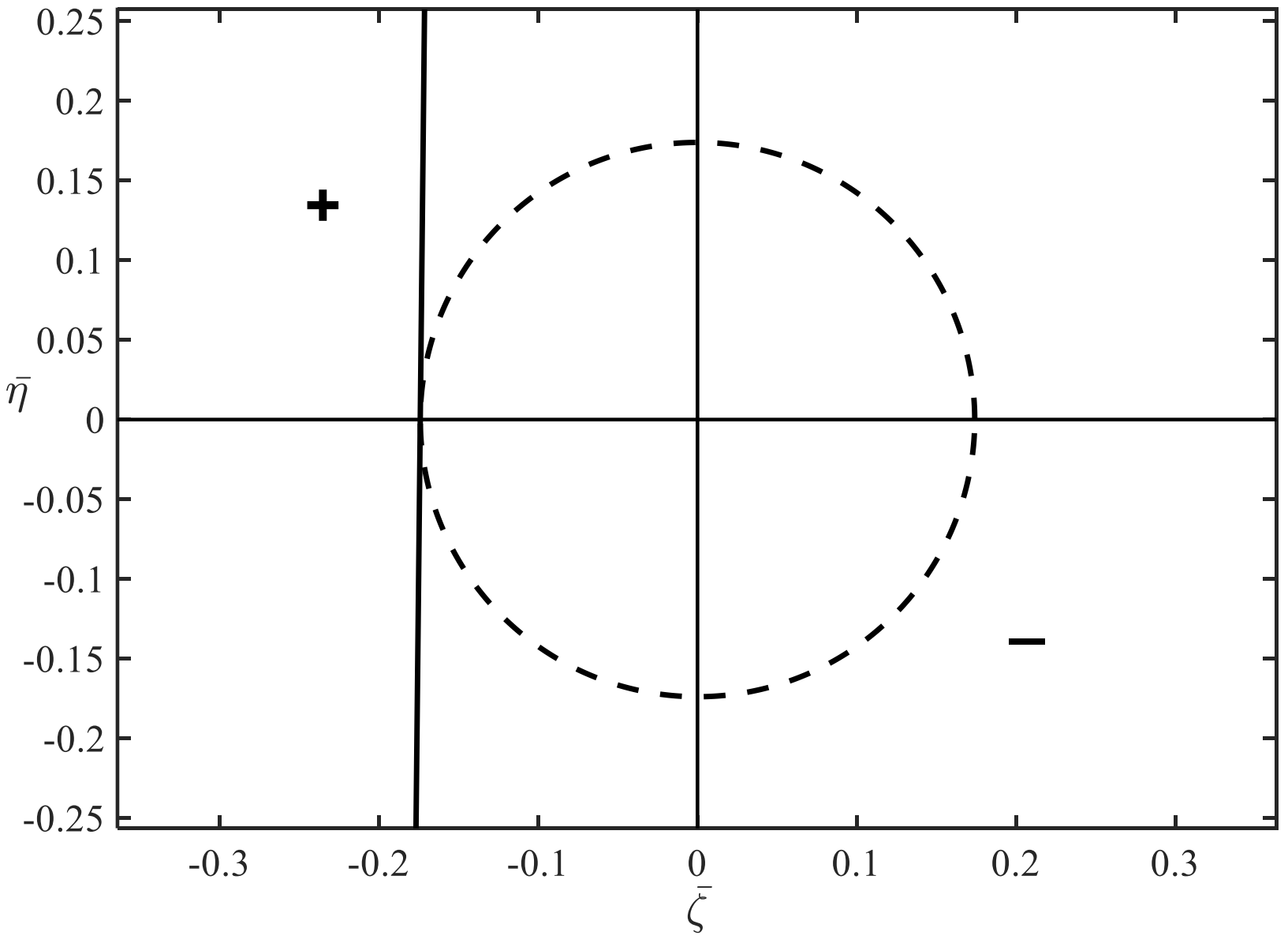

**Fig. 2(b).** Real and Imaginary parts of normalized voltage applied on the spherical object and the repulsive (positive,+) and attractive (negative,-) force zones for non-dimensional frequency of $k_f a = 20$.The black line shows the zero-radiation force state and the dashed circle shows the tangent circle, which its radius shows the minimum required voltage for zero-force state.

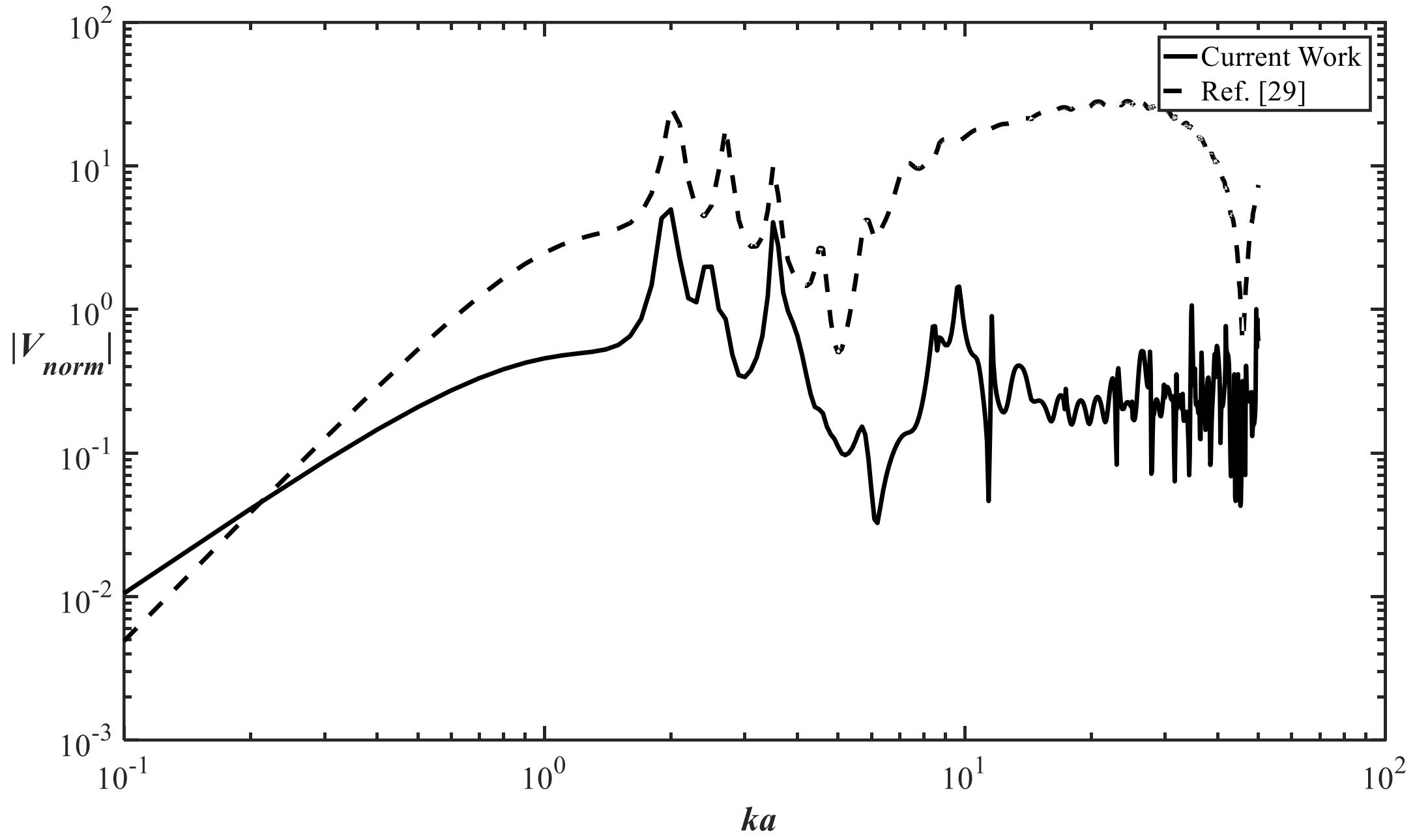

**Fig. 3(a).** Amplitude of non-uniform voltage, $|V|$, required to cancel the acoustic radiation force in *Z*-direction on carrier with outer radius of 1 mm which is insonified by a progressive plane wave with the incident pressure amplitude of 100 Pa versus non-dimensional frequency, $k_f a$, in logarithmic scale and the comparison with the case of breathing mode voltage implementation, introduced in Ref. [32].

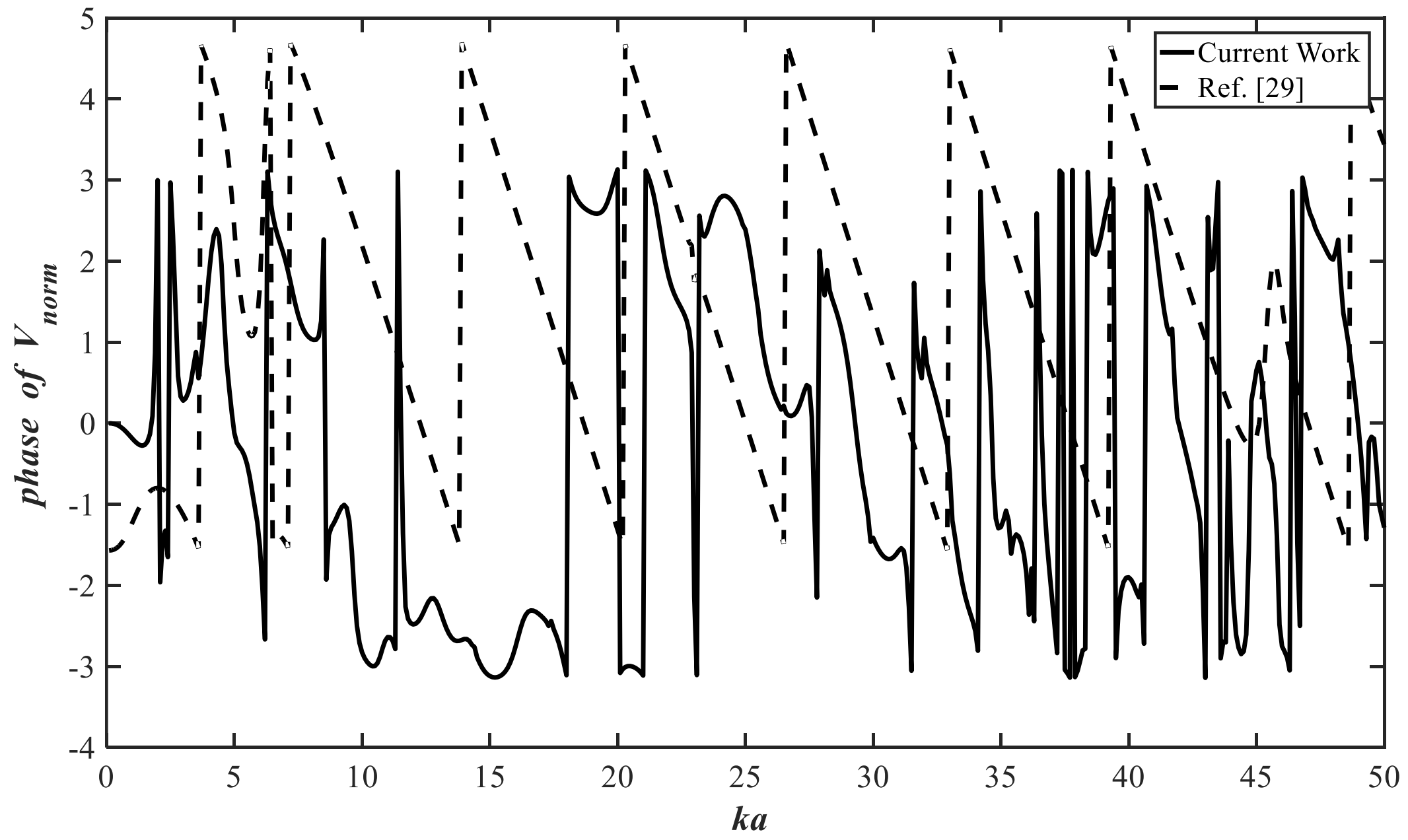

**Fig. 3(b).** Phase of non-uniform voltage, $|V|$, required to cancel the acoustic radiation force in $Z$-direction on carrier with outer radius of 1 mm which is insonified by a progressive plane wave with the incident pressure amplitude of 100 Pa versus non-dimensional frequency, $k_f a$, in logarithmic scale and the comparison with the case of breathing mode voltage implementation, introduced in Ref. [32].

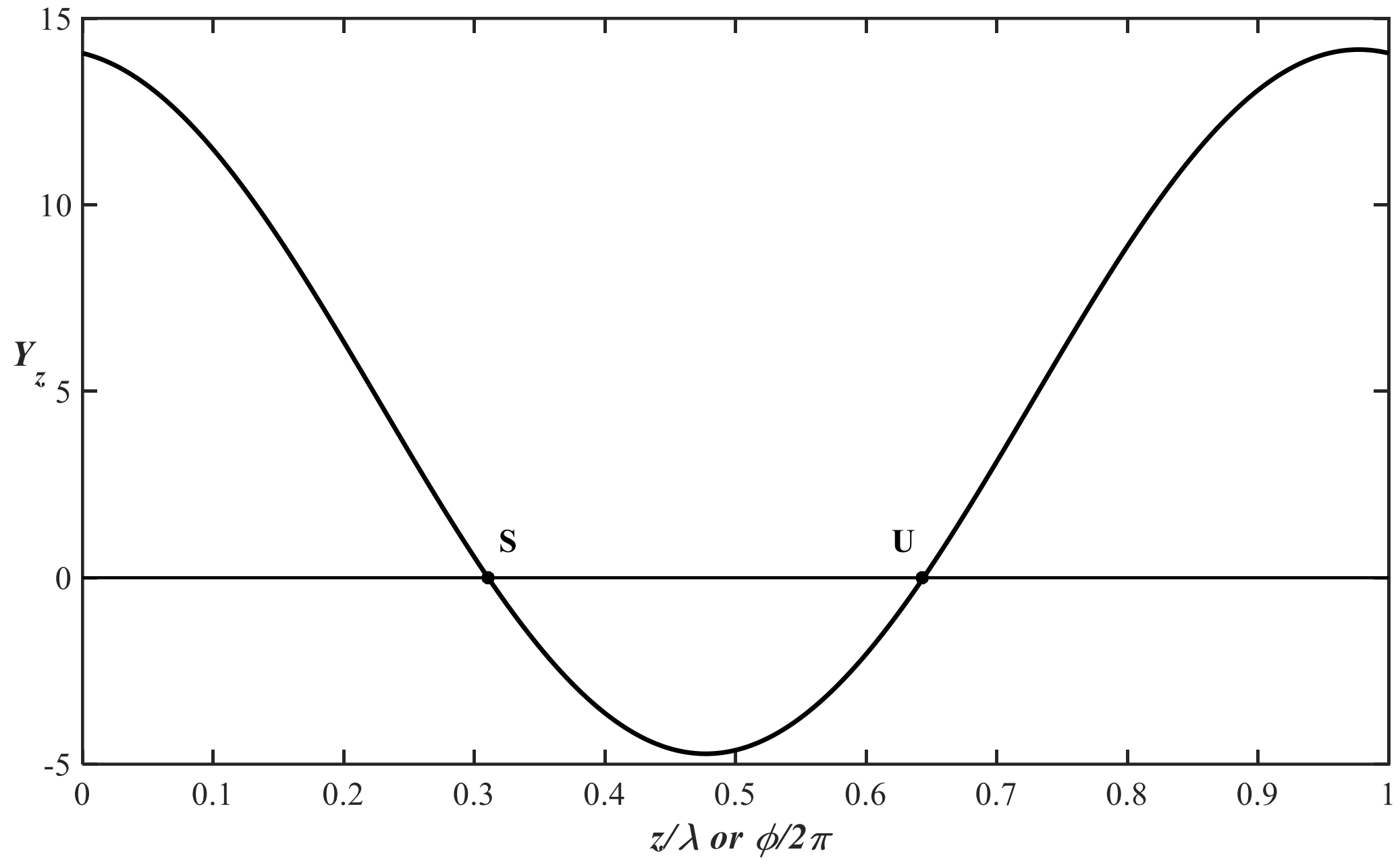

**Fig. 4(a).** The non-dimensional radiation force as a function of distance, $z/\lambda$, or phase difference, $\tan^{-1}(\bar{\eta}/\bar{\varsigma})$, at $k_f a = 2$ and $\alpha = 0^{\circ}$. *S* and *U* indicate the stable and unstable equilibrium states.

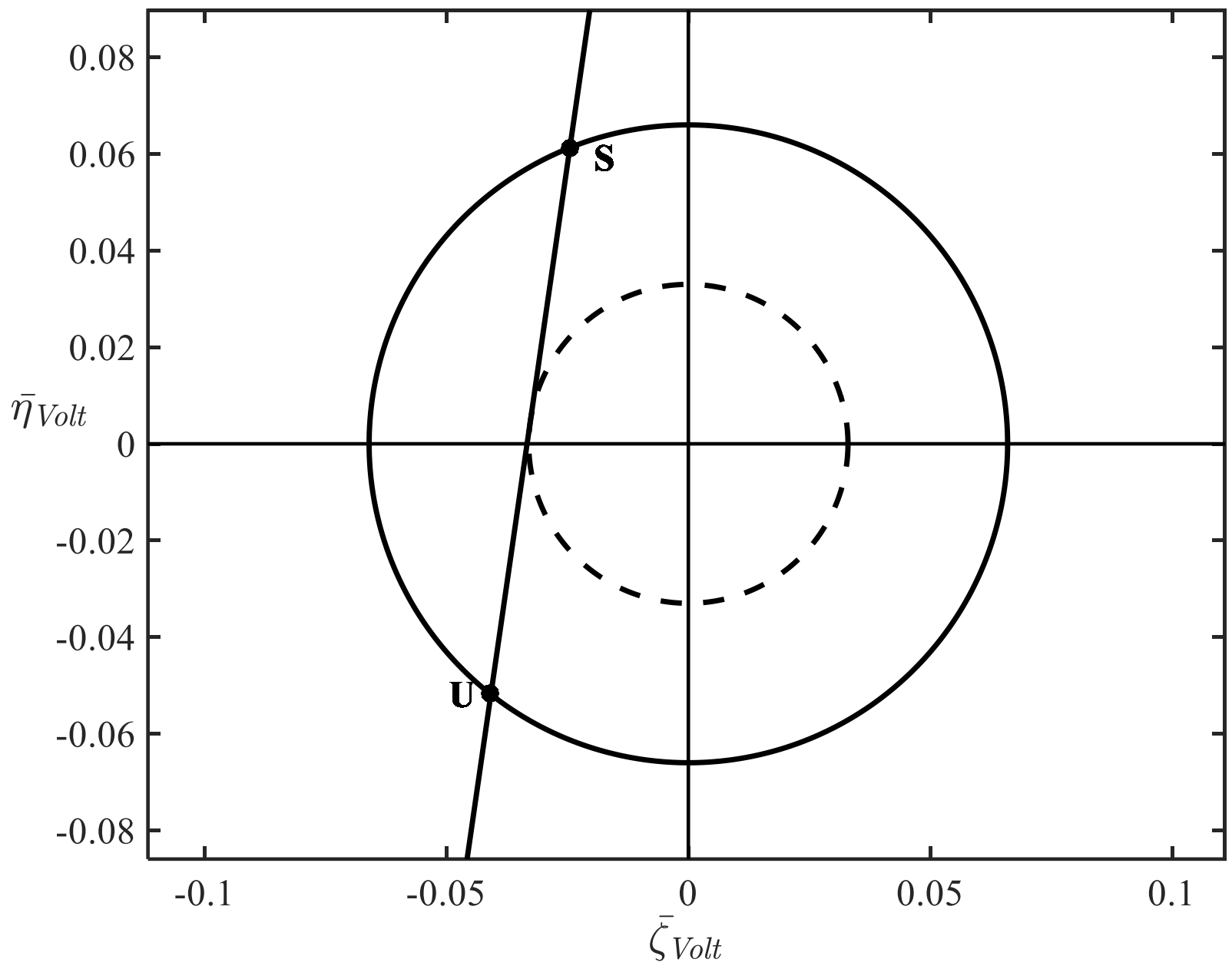

**Fig. 4(b).** The Nyquist (Real and Imaginary) plot of voltage for $R = 2$, at $k_f a = 2$ and $\alpha = 0^{\circ}$. *S* and *U* indicate the stable and unstable equilibrium states.

Considering the aim of this paper to evaluate the possibility of steering a spherical active carrier makes us to review that in this technique, the controllable parameter is the polar orientation, $\alpha$ , of the divisor plane installed on the inner layer of the sphere which its rotation is independent of the rotation of the whole sphere. In the rest of the paper, it is assumed that following the successful trapping strategy of active carriers [10], the carrier is trapped at the stable zero-force state. Now, the question is the 3D maneuverability of the carrier: Is it possible to induce a net radiation force on the active carrier along any desired direction, just by altering the orientation of the divisor plane?

Fig. 5 shows the direction of the net radiation force, $\gamma$, as a function of the orientation of divisor plane, $\alpha$ , at selected frequency, $k_f a = 2$. It is seen that the whole range of $-180^o < \gamma < 180^o$ is covered. The same results has been obtained for a great frequency range, specially the frequency range of $k_f a < 10$ which maps the practical frequency range of $\omega / (2\pi) < 10 MHz$ . The special divisor angles of $\alpha = \pm 133.5^\circ$, indicate two special interesting cases of $\gamma = \mp 90^\circ$, as desired cases I and II which the carrier moves downward and upward, respectively and it is perpendicular to the incident beam direction.

Figure 6 shows the amplitude of the net radiation force (non-dimensional) as a function of the orientation of divisor plane, $\alpha$ . Figure 7 depicts the radiation force components, $\Gamma_X$ and $\Gamma_Z$ , as functions of the orientation of divisor plane, $\alpha$ . The variation of force components with respect to the divisor angle is smooth and fortunately, no sign of singularity or jump or drop is seen. Figure 8(a) and (b) illustrate the schematic configurations of the two special desired cases.

The concern of rotational stability should be investigated and addressed. Figure 9 illustrate the acoustic radiation torque as a function of the orientation of divisor plane, $\alpha$ . It is seen that the induced radiation torque is at its maximum state for $\alpha = \pm 90^\circ$. More importantly, it is observed that the carrier at the symmetric configuration (i.e., $\alpha = 0^\circ$) is rotationally stable and any deviation will be compensated by the stabilizing radiation torque (i.e., $\tau_y > 0 \; for \, \alpha > 0$ and $\tau_y < 0 \; for \, \alpha < 0$). For other symmetric configuration at $\alpha = \pm 180^\circ$, the carrier is rotationally unstable (i.e., $\tau_y < 0 \; for \, \alpha > 180^\circ$ and $\tau_y > 0 \; for \, \alpha < 180^\circ$). It is clear that the rotational unstability enforce the carrier to be oriented along its rotationally stable orientation. Therefore, no serious concern about the rotational stability in the case of manipulation direction parallel to the navigation wave field occurs; but, the theoretically maximum force and velocity of manipulation, obtained in previous paragraph, is not achievable.

Returning to Fig. 5, a discontinuity at $\alpha = 0^\circ$ is observed for which, the angle of net force direction jump to near $\gamma = \pm 90^o$. From Fig. 7, it is seen that both force components are zero at $\alpha = 0^\circ$, but as the angle deviates from zero, the amplitude of $\Gamma_X$ is increased much rapidly with respect to the amplitude of $\Gamma_Z$; therefore, the jump occurs for $\gamma(\alpha = 0^\circ) = 0^\circ$, $\gamma(\alpha \to 0^\circ) = \lim_{\alpha \to \pm 0^\circ} \tan^{-1}(\Gamma_X / \Gamma_Z) = \lim_{\alpha \to \pm 0^\circ} \tan^{-1}(\varepsilon / 0) = \pm 90^\circ$. Considering the order of amplitude of $\Gamma_X \sim O(10^{-2})$ for $\alpha < \pm 3^o$, and the stabilizing radiation torque at $\alpha = 0^\circ$, the scenario of unwanted transverse deviation of carrier is rejected.

One challenge of the proposed 3D translational manipulation technique is the non-zero torque which may deviate the carrier from the desired direction. For the special interesting case of perpendicular motion of the carrier with respect to the incident wave direction (i.e., $\gamma = \pm 90^o$, $\alpha = \mp 133.5^o$), an estimation of the time and the displacement and the torque-generated deviation can be done for the desired case 1 or 2 at $k_f a = 2$. Assuming Stokes pseudo steady-state condition [46], we have the torque balance as $N_Y + 8\pi\mu a^3 \dot{\alpha} = 0$, and the force balance as $F_x - 6\pi\mu a U_x = 0$, where $\mu$ is dynamics viscosity of host medium, $\dot{\alpha} = d\alpha / dt$ is the angular velocity of carrier, $U_x$ is the linear velocity of carrier along *X*-axis and $a$ is the outer radius of the object, we can deduce an estimated value for the order of angular and linear velocity. By considering $a \sim O(10^{-3})\,\mathrm{m}$, $\mu \sim O(10^{-4})\,\mathrm{Pa.s}$, $N_Y = \pi a^3 (\rho_0 k^2 \phi_0{}^2 / 2)\tau_Y \sim O(10^{-15})\,\mathrm{N.m}$, $F_X = \Gamma_X \pi a^2 p_0^2 / (2\rho_0 c_0^2) \sim O(10^{-11})\,\mathrm{N}$, the order of obtained linear and angular velocity will be $\sim O(10^{-5})\,\mathrm{m/s}$ and $\sim O(10^{-4})\,\mathrm{rad/s}$. To obtain an obvious physical vision, we can compare the amount of rotation of the sphere when it has moved a distance equal to its radius, perpendicular to the incident beam direction at the desired cases I or II. The order of time required for the carrier to go along for 1 mm is about $\sim O(10^2)\,\mathrm{s}$ and based on the angular velocity, the carrier will rotate about $0.57^\circ$ which is so negligible.

Considering the above feasibility study in one hand and on the other hand, this simple fact that any dislocation of the carrier perpendicular to the direction of beam does not change the phase difference between the beam and the voltage state, the strategy of 3D steering/manipulation may be as below:

Without any conservation about the position of the carrier with respect to the incident beam field, just by setting the operational frequency of both transducers (i.e., active parts of carrier and incident beam generator), the carrier will be trapped on the nearest stable equilibrium position with a dislocation shorter than one wavelength of the beam. Moreover, the carrier will be aligned in its symmetric state (i.e., $\alpha = 0^\circ$) with respect to the beam direction, due to the induction of a stabilizer radiation torque. The degree of manipulability along the beam direction is achievable by just slow phase variation of the incident acoustic field, as well as what happens in standing wave manipulation strategy [44,45]. For transverse steering, depending on the frequency of operation, a specific orientation of the divisor plane always exists which the net radiation force will be perpendicular to the beam direction and the transverse motion is possible with the same agility of manipulation along the beam direction. Therefore, in any desired *XZ*- plane (i.e., *Z*-axis along the beam direction and *X*-axis designates the transverse direction), two degree of manipulability exists which in 3D space, means three degree of manipulability (i.e., the *X*-axis as well as the *X'*-axis can be rotated about the *Z*-axis and the full cover of polar coordinate perpendicular to the *Z*-axis is achievable).

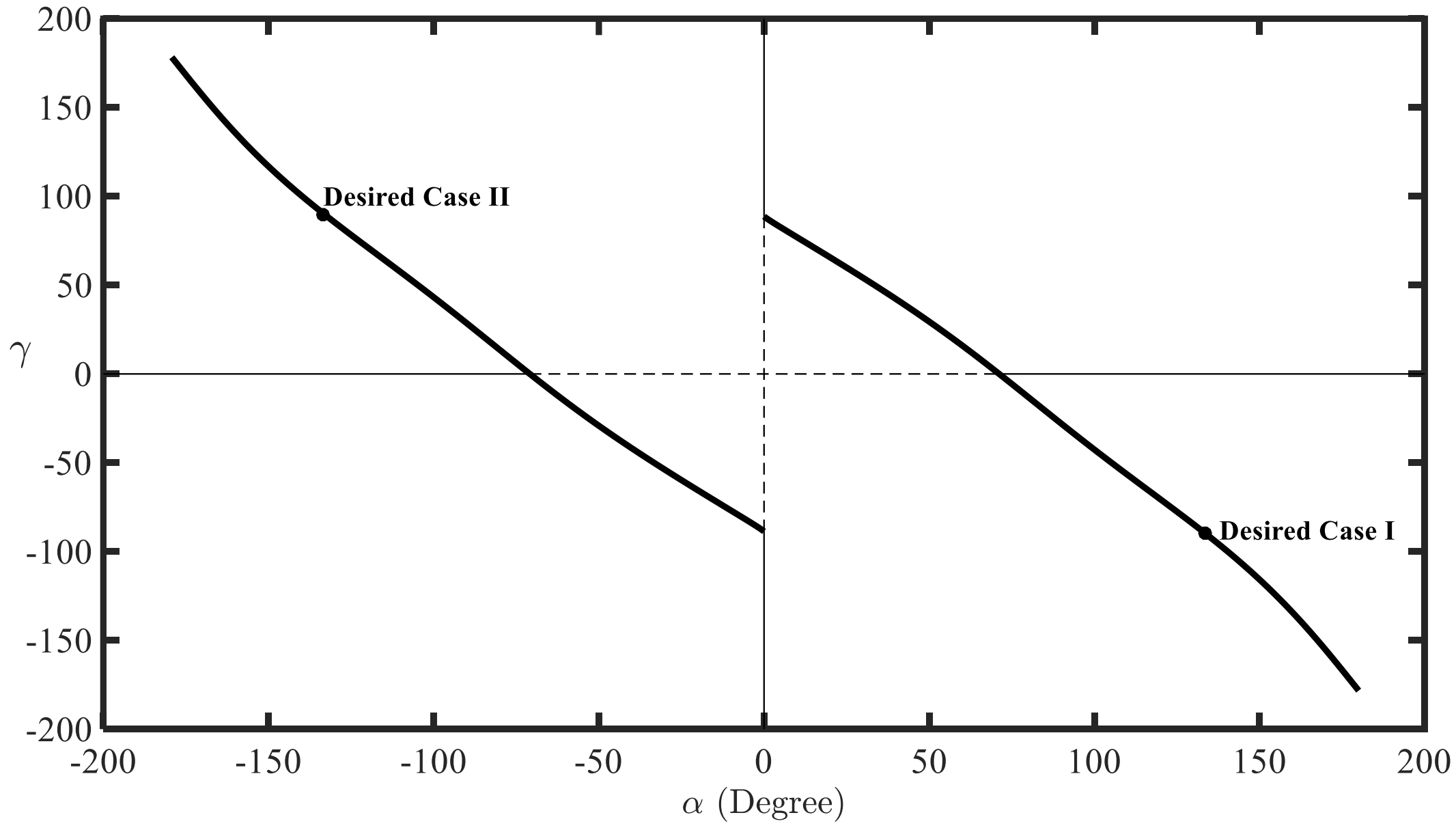

**Fig. 5.** The direction of the net radiation force, $\gamma$, as a function of the orientation of divisor plane, $\alpha$, at selected frequency, $k_f a = 2$.

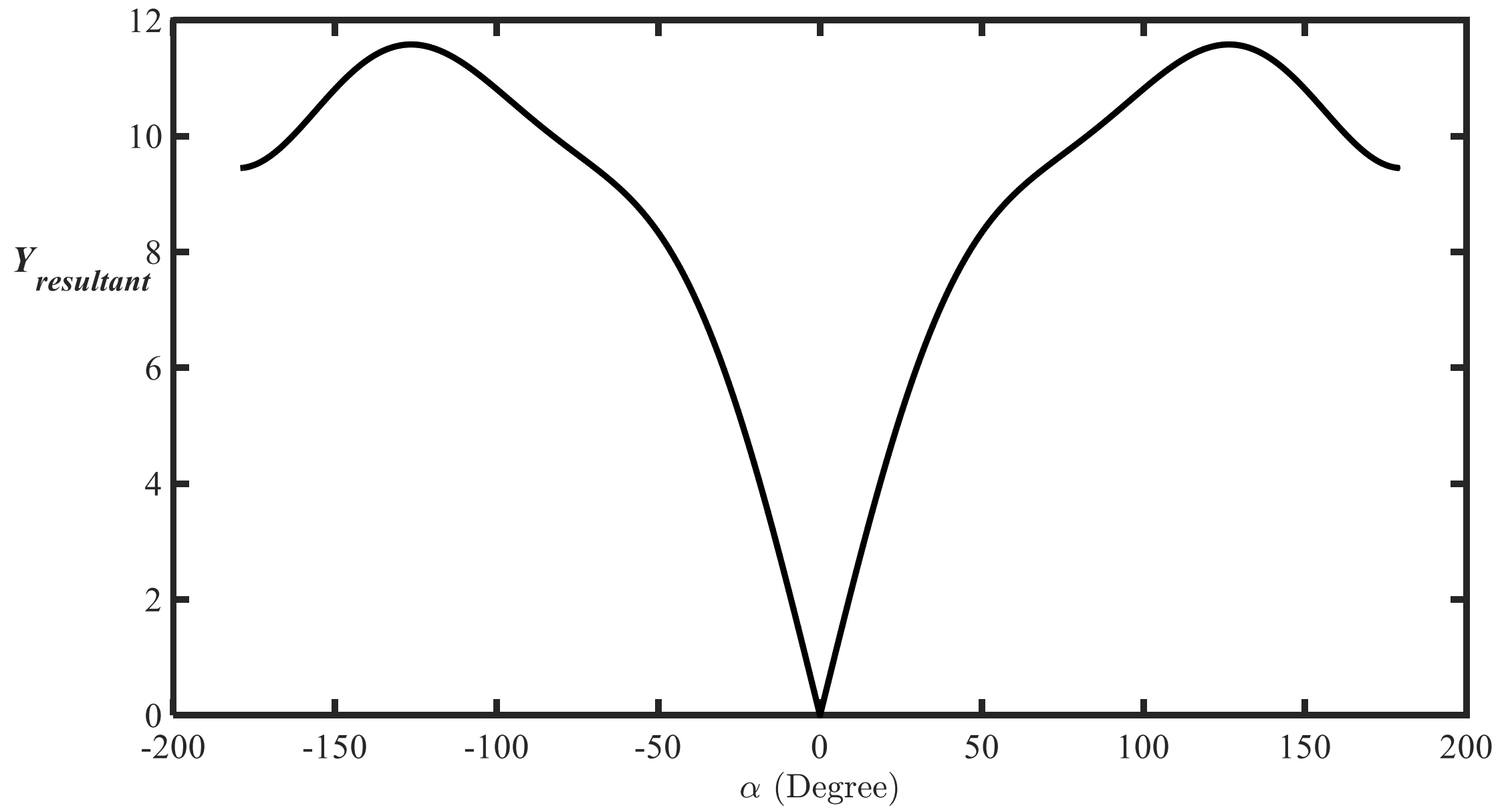

**Fig. 6.** Amplitude of net non-dimensional acoustic radiation force, $\left(\Gamma_X^2+\Gamma_Z^2\right)^{1/2}$, versus the angle of divisor plane, $\alpha$, at $k_f a=2$.

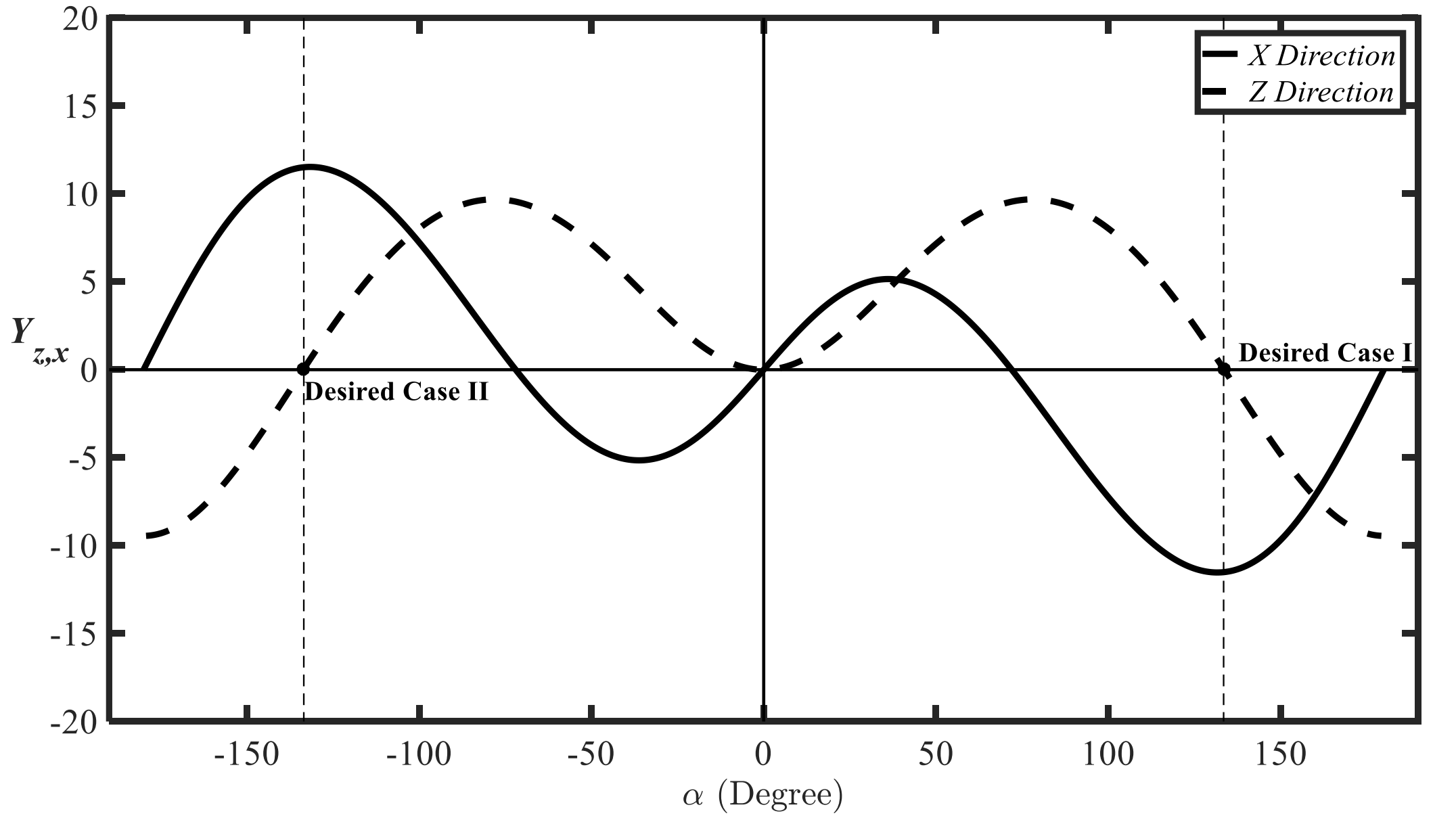

**Fig. 7.** Non-dimensional acoustic radiation force components, $\Gamma_X$ (solid line), and $\Gamma_Z$ (dashed line) at $k_f a=2$, at stable equilibrium state, as a function of the polar orientation of divisor plane, $\alpha$.

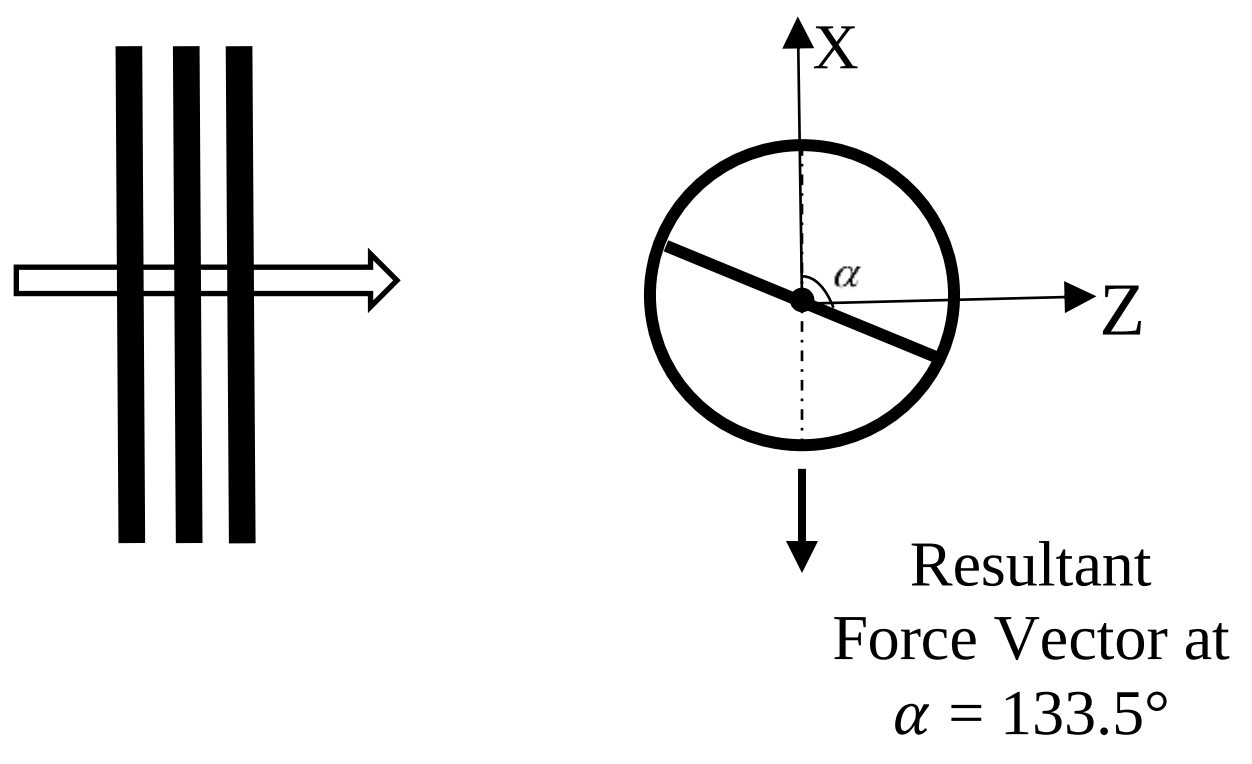

**Fig. 8(a).** Desired case 1: The resultant vector of force is merely in *X* direction and its polar angle $\gamma = -90^{\circ}$. This phenomenon occurs at $\alpha = +133.5^{\circ}$

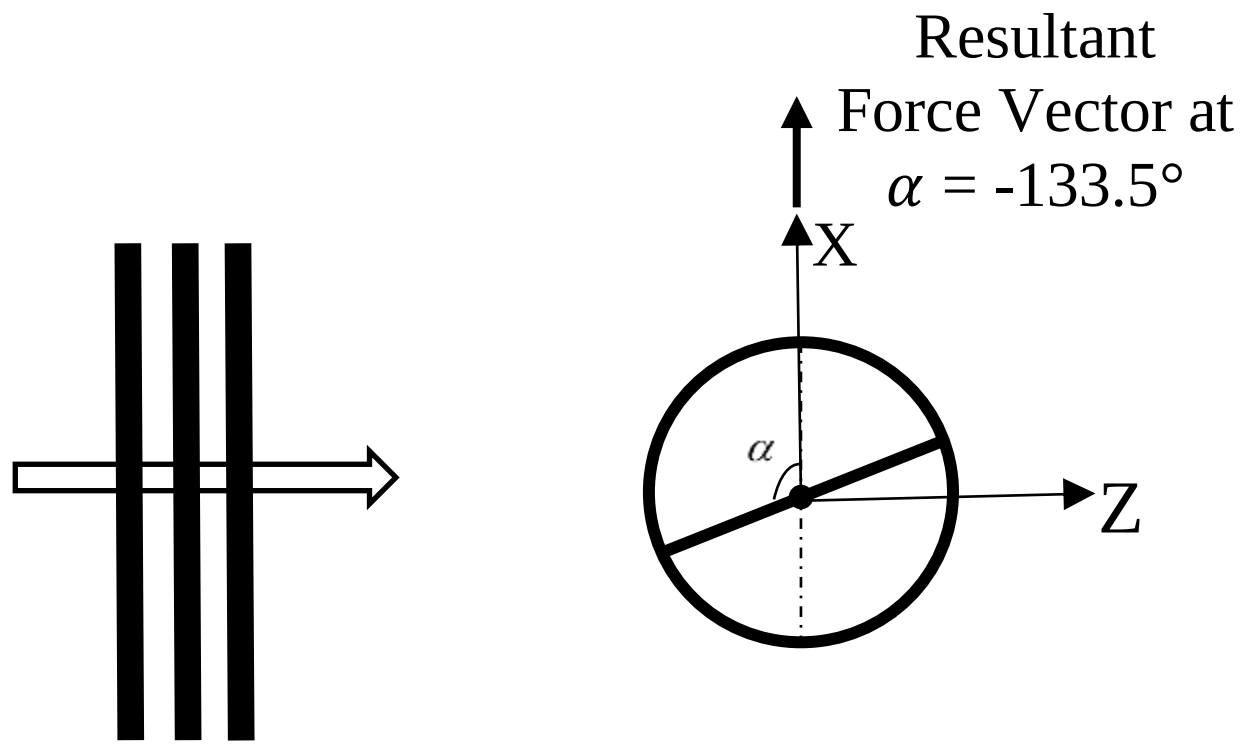

**Fig. 8(b).** Desired case 2: The resultant vector of force is merely in *X* direction and its polar angle $\gamma = +90^{\circ}$. This phenomenon occurs at $\alpha = -133.5^{\circ}$

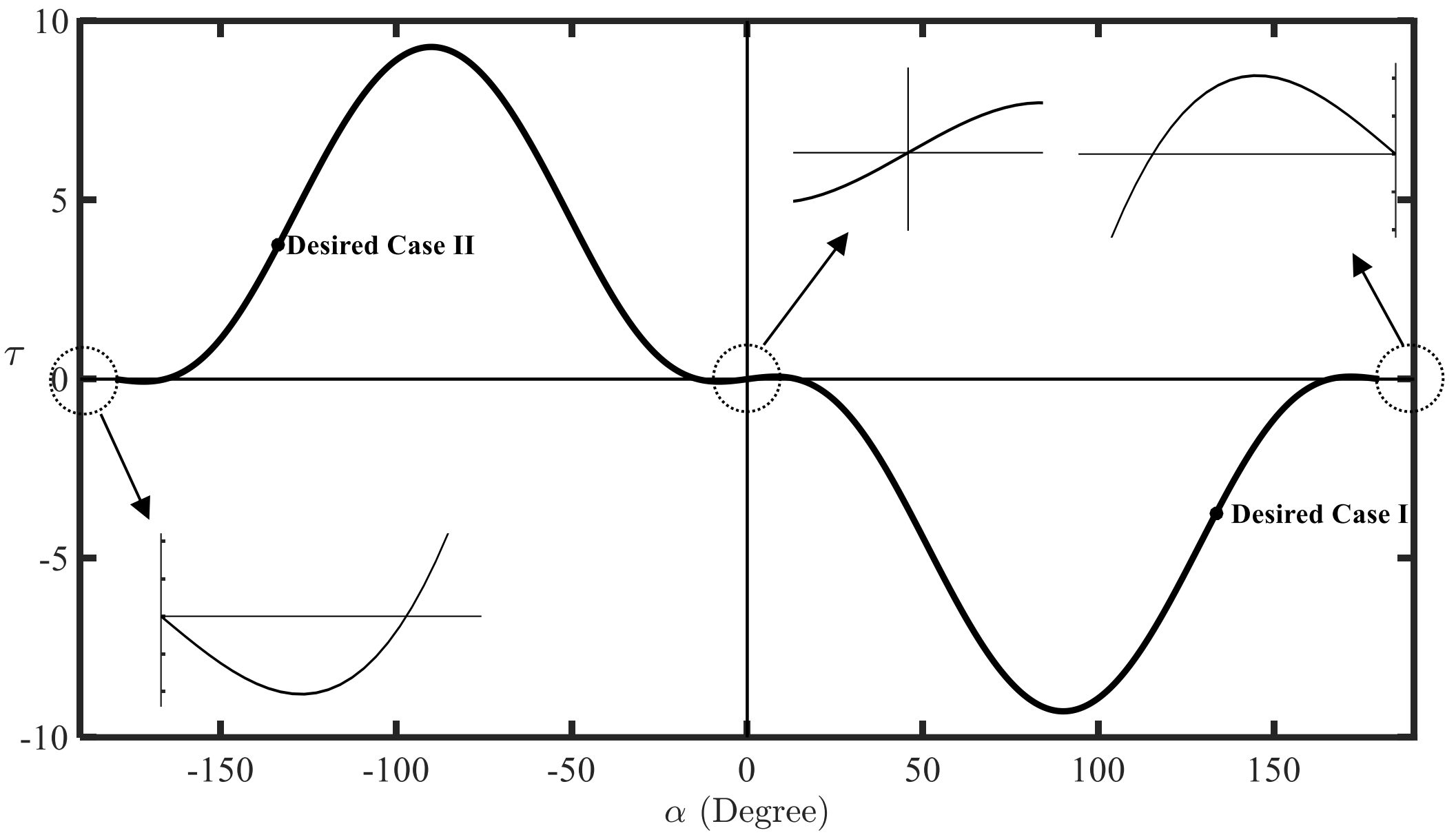

**Fig. 9.** Non-dimensional acoustic radiation torque about $Y$-axis at $k_f a = 2$, at stable equilibrium state, as a function of the polar orientation of divisor plane, $\alpha$ . The figure shows the rotational stability for $\alpha = 0^{\circ}$ and instability for $\alpha = 180^{\circ}$ .

## 5. Conclusion

In the present work, 3D manipulability and the possibility of steering of the active carriers has been analyzed. It is assumed that the introduced active carrier is equipped with internal bipolar configuration of piezoelectric driver which its divisor plane between its two activated semi-sphere parts may have any desired orientation. It has been shown that all three degrees of translational manipulability is achievable. The feasibility study on the required voltage, the achieved force and the velocity of manipulation prove some merits of the active carriers as an asset for advanced material/agent/drug delivery systems in small scale tasks. Some feasibility challenges about the stability of steering and manipulation process are considered and discussed. It is clear that the present work should be considered as an elementary step toward manipulability of active carriers and the raised complexities due to the design and fabrication of the proposed theoretical configuration need to be solved in future works.

## Appendix

$$\Omega_n^m = \begin{pmatrix} S_n^m(1,3) & S_n^m(1,4) & S_n^m(1,5) & \frac{-\omega}{C_{44}}\rho_f\varphi_0 i^{n+1}Y_{n,m}^*(\alpha,0)h_n^{(1)}(k_f a) \\ S_n^m(2,3) & S_n^m(2,4) & S_n^m(2,5) & 0 \\ S_n^m(4,3) & S_n^m(4,4) & S_n^m(4,5) & \frac{-\varphi_0}{a\omega i}k_f i^n Y_{n,m}^*(\alpha,0)h_n^{(1)\prime}(k_f a) \\ T_n^m(6,3) & T_n^m(6,4) & T_n^m(6,5) & 0 \end{pmatrix},$$

$$X_n^m = \begin{pmatrix} G_n^m \\ W_n^m \\ \delta_n^m \\ A_n^m \end{pmatrix},$$

$$T_n^m = \begin{pmatrix} \frac{\omega}{C_{44}}Y_{n,m}^*(\alpha,0)\rho_f\varphi_0 i^{n+1}j_n(k_f a) \\ 0 \\ \frac{\varphi_0}{i\omega a}k_f i^n Y_{n,m}^* j_n{}'(k_f a) \\ \Phi_n \end{pmatrix}.$$

$$Z_{n,m}^1 = \frac{\begin{vmatrix} S_n^m(1,3) & S_n^m(1,4) & S_n^m(1,5) & \frac{\omega}{C_{44}}Y_{n,m}^*(\alpha,0)\rho_f\varphi_0 i^{n+1}j_n(k_f a) \\ S_n^m(2,3) & S_n^m(2,4) & S_n^m(2,5) & 0 \\ S_n^m(4,3) & S_n^m(4,4) & S_n^m(4,5) & \frac{\varphi_0}{i\omega a}k_f i^n Y_{n,m}^* j_n{}'(k_f a) \\ T_n^m(6,3) & T_n^m(6,4) & T_n^m(6,5) & 0 \end{vmatrix}}{\begin{vmatrix} S_n^m(1,3) & S_n^m(1,4) & S_n^m(1,5) & \frac{-\omega}{C_{44}}\rho_f\varphi_0 i^{n+1}Y_{n,m}^*(\alpha,0)h_n^{(1)}(k_f a) \\ S_n^m(2,3) & S_n^m(2,4) & S_n^m(2,5) & 0 \\ S_n^m(4,3) & S_n^m(4,4) & S_n^m(4,5) & \frac{-\varphi_0}{a\omega i}k_f i^n Y_{n,m}^*(\alpha,0)h_n^{(1)\prime}(k_f a) \\ T_n^m(6,3) & T_n^m(6,4) & T_n^m(6,5) & 0 \end{vmatrix}}, n > 0$$

$$Z_{n,m}^{2}=\frac{\begin{vmatrix} S_n^m(1,3) & S_n^m(1,4) & S_n^m(1,5) & 0\\ S_n^m(2,3) & S_n^m(2,4) & S_n^m(2,5) & 0\\ S_n^m(4,3) & S_n^m(4,4) & S_n^m(4,5) & 0\\ T_n^m(6,3) & T_n^m(6,4) & T_n^m(6,5) & 1\end{vmatrix}}{\begin{vmatrix} S_n^m(1,3) & S_n^m(1,4) & S_n^m(1,5) & \frac{-\omega}{C_{44}}\rho_f\varphi_0 i^{n+1}Y_{n,m}^{*}(\alpha,0)h_n^{(1)}(k_f a)\\ S_n^m(2,3) & S_n^m(2,4) & S_n^m(2,5) & 0\\ S_n^m(4,3) & S_n^m(4,4) & S_n^m(4,5) & \frac{-\varphi_0}{a\omega i}k_f i^{n}Y_{n,m}^{*}(\alpha,0)h_n^{(1)\prime}(k_f a)\\ T_n^m(6,3) & T_n^m(6,4) & T_n^m(6,5) & 0\end{vmatrix}},n>0$$

where, $S_n^m(i,j)$ are the elements of the global structural transfer matrix, $G_n^m$ and $W_n^m$ are displacement functions which are indicated in our previous work [32]. It is also notable to mention that the expression for $Z_{n,m}^2$ and $Z_{n,m}^1$ for $n=0$ is like the one for $n>0$ with the difference that the second rows of the matrices should be eliminated.